\documentclass[a4paper,12pt]{article}

\renewcommand\epsilon{\varepsilon}
\def\bea{\begin{eqnarray}}
\def\eea{\end{eqnarray}}
\def\nn{\nonumber}
\makeatletter
\def\vereq#1#2{
\lower3pt\vbox{\baselineskip1.5pt \lineskip1.5pt
\ialign{$\m@th#1\hfill##\hfil$\crcr#2\crcr\sim\crcr}}}
\makeatother

\begin{document}
\begin{titlepage}
\begin{center}
\hfill    CERN-TH/2001-170\\
\hfill HUPD-0109\\
~{} \hfill hep-ph/0107164\\
\vskip 1cm

{\large \bf A Bridge between CP violation
at Low Energies and Leptogenesis}

\vskip 1cm

Gustavo C. Branco\footnote{gbranco@thwgs.cern.ch and 
gbranco@cfif.ist.utl.pt} \footnote{On leave of absence from 
Departamento de F\'\i sica,
Instituto Superior T\' ecnico, Av. Rovisco Pais, P-1049-001, Lisboa,
Portugal.} 
Takuya Morozumi\footnote{takuya.morozumi@cern.ch and 
morozumi@theo.phys.sci.hiroshima-u.ac.jp}
\footnote{On leave of absence from
Graduate School of Science, Hiroshima University 
1-3-1 Kagamiyama, Higashi Hiroshima - 739-8526, Japan}
B. M. Nobre\footnote{bnobre@cfif.ist.utl.pt}
\footnote{Departamento de F\'\i sica,
Instituto Superior T\' ecnico, Av. Rovisco Pais, P-1049-001, Lisboa,
Portugal.} 
and 
M. N. Rebelo
\footnote{mrebelo@thwgs.cern.ch and
rebelo@cfif.ist.utl.pt}\ $^ \dagger$
\vskip 0.05in

{\em Theory Division, CERN, CH-1211 Geneva 23, Switzerland.}
\end{center}

\vskip 3cm

\begin{abstract}
We discuss the possibility of relating
the size and sign of the observed baryon asymmetry 
of the universe to
CP violation observable at low energies,
in a framework where the observed baryon asymmetry 
is produced by leptogenesis
through the out of the equilibrium decay of heavy 
Majorana neutrinos. We identify the CP 
violating phases entering in leptogenesis as well 
as those relevant for CP violation at low energies
in the minimal seesaw model. We
show that although in general there is no relation
between these two sets of phases, there are specific frameworks
in which such a connection 
may be established and we give a specific
grand unification inspired example where 
such a connection does exist. We construct
weak-basis invariants related to CP violation
responsible for leptogenesis, as well as those relevant 
for CP violation at low energies.
\end{abstract}

\end{titlepage}

\newpage 
\section{Introduction}
In the Standard Model (SM) neutrinos are massless and there is no
mixing or CP violation in the leptonic sector. However, in any
extension of the SM which incorporates neutrino masses and mixing,
CP violation naturally arises also in the leptonic sector and can
in principle be measured through neutrino oscillations \cite{osc}.
CP violation in the leptonic sector can have profound cosmological
implications, playing a crucial r\^ ole in the generation of the
observed baryon number asymmetry of the universe (BAU). 
Indeed, one of the most
plausible scenarios for the generation of BAU is through the out-of-
equilibrium decay of heavy Majorana neutrinos \cite{Fukugita:1986hr}. 
This mechanism has been 
studied in detail by several groups \cite{buch} and
it has been shown that the
observed BAU, $n_B/s \sim 10^{-10}$, can be obtained in the above
scenario, without any fine-tuning of parameters.
In this paper, we address the question of whether it is possible 
to establish a connection between CP breaking necessary to 
generate leptogenesis and CP violation at low energies. 
More specifically, assuming that 
baryogenesis is achieved through leptogenesis, can one infer the 
strength of CP violation at low energies from the size
and sign of the observed BAU?
We show that without any further assumption
about the flavour structure of the leptonic mass matrices and/or the
mechanism of CP violation, in general one cannot establish a direct 
connection,between the strength of CP violation at high energies
(required for leptogenesis) and that observed at low energies. 
However, we shall point out that there are special scenarios
where such a connection may be established. In particular, we describe 
a scenario inspired on grand unified theories (GUTs),
where such a connection arises.
Mixing and CP violation with Majorana neutrinos is often described in the
decoupling limit via the $3 \times 3$ Maki-Nakagawa-Sakata 
matrix \cite{Maki:1962mu}.   
Since for leptogenesis, we have to analyze decays of
heavy neutrinos, we need to consider the full $3 \times 6$
mixing matrix appearing in the weak charged current, connecting the  
charged leptons to the three light neutrinos and the three heavy neutrinos.
The number of independent CP violating phases 
was identified for the general case in Ref.\cite{Branco:1986gr} 
and for a special case in Ref.\cite{Endoh:2001hc}.
In this paper, we identify the CP violating phases both in an
appropriately chosen weak-basis (WB) and in the mass eigenstate basis.
Furthermore, we construct WB invariants which must vanish if CP
invariance holds. Non-vanishing of any of these WB invariants signals CP
violation. We identify the WB invariants relevant
to leptogenesis and those associated to CP violation at low energies.
The interest of these WB invariants stems from the fact that they
can be evaluated in any WB and are thus particularly suited to
the analysis of specific ans\" atze for charged lepton and neutrino
mass matrices. For any given ansatz, in order to analyze its potential
for
leptogenesis or whether it leads to CP violation in neutrino
oscillations,
one can simply compute the appropriate WB invariant.
Most of our analysis does not depend on the origin of CP
violation, namely on whether CP is explicitly broken at the
Lagrangian level or spontaneously broken. However, we present
a simple extension of the SM where CP is spontaneously broken 
by a single phase of the vacuum expectation value of a 
complex scalar, singlet 
under $SU(3)_{c} \times SU(2)_{L} \times U(1)$.
It is shown that this phase is sufficient to produce both 
the CP violation necessary for baryogenesis and the 
CP violation observable at low energies.
\section{ A minimal extension of the SM}
\setcounter{equation}{0}
Let us consider a minimal extension of the SM which consists of adding
to the standard spectrum one right-handed neutrino per generation.
After spontaneous gauge symmetry breaking, the following leptonic mass
terms
can be written:
\bea
{\cal L}_m  &=& -[\overline{{\nu}_{L}^0} m_D \nu_{R}^0 +
\frac{1}{2} \nu_{R}^{0T} C M_R \nu_{R}^0+
\overline{l_L^0} m_l l_R^0] + h. c. = \nonumber \\
&=& - [\frac{1}{2}  n_{L}^{T} C {\cal M}^* n_L +
\overline{l_L^0} m_l l_R^0 ] + h. c.
\label{lm}
\eea
where $m_D$, $M_R$ and $m_l$ denote the neutrino Dirac mass matrix,
the right-handed neutrino Majorana mass matrix and the charged
lepton mass matrix, respectively, and
$n_L = ({\nu}_{L}^0, {(\nu_R^0)}^c)$.
The right-handed neutrino Majorana mass term is
SU(2) $\times$ U(1) invariant, consequently it can have a value 
much above the scale $v$ of the electroweak symmetry breaking, 
thus leading, through the seesaw mechanism \cite{see},
to naturally small left-handed Majorana neutrino masses,
of order $\frac{{m_D}^2}{M_R}$.
It is convenient to determine the nature of the various CP
violating phases, both in a weak-basis (WB), where all gauge
currents are real and flavour diagonal, and in a mass-eigenstate basis, 
where fermion mass terms are real, diagonal but there is
non-trivial
flavour mixing in the charged currents.
\subsection{ CP violating phases in a weak-basis}
In order to study CP violation in a WB, let us first note that the
most general CP transformation which leaves the gauge
interaction invariant is:
\bea
{\rm CP} l_L ({\rm CP})^{\dagger}&=&U \gamma^0 {\rm C} \overline{l_L}^T
\quad
{\rm CP} l_R({\rm CP})^{\dagger}=V \gamma^0 {\rm C} \overline{l_R}^T
\nonumber \\
{\rm CP} \nu_L ({\rm CP})^{\dagger}&=&U \gamma^0 {\rm C}
\overline{\nu_L}^T \quad
{\rm CP} \nu_R ({\rm CP})^{\dagger}=W \gamma^0 {\rm C} \overline{\nu_R}^T
\label{cp}
\eea
where U, V, W are unitary matrices acting in flavour space
and where for notation simplicity we have dropped here the
superscript 0 in the fermion fields. .
Invariance of the mass terms under the above CP transformation,
requires that the following relations have to be satisfied:
\bea
W^T M_R W &=&-M_R^*  \label{cpM} \\
U^{\dagger} m_D W&=& {m_D}^*  \label{cpm} \\
U^{\dagger} m_l V&=& {m_l}^* \label{cpml}
\eea
In order to analyze the implications of the above conditions,
it is convenient to choose the WB where both $m_l$, $M_R$, are real
diagonal. In this basis, W is then constrained by
Eq.~(\ref{cpM})  to be of the form
\bea
W={\rm diag.} \left(\exp(i\alpha_1), \exp(i\alpha_2),... \exp(i\alpha_n)
\right) \label{expw}
\eea
where n denotes the number of generations and the $\alpha_i$ have to
satisfy:
\bea
\alpha_i=(2 p_i +1) \frac{\pi}{2} \label{ais}
\eea
with $p_i$ integer numbers. Multiplying Eq.~(\ref{cpml}) by its Hermitian
conjugate,
and  taking into account that we are working on a WB where $m_l$ is
real diagonal,
one concludes that U has to be of the form:
\bea
U={\rm diag.} \left(\exp(i\beta_1), \exp(i\beta_2),... \exp(i\beta_n)
\right) \label{expu}
\eea
where $\beta_i$ are arbitrary phases. From Eqs.~(\ref{cpm}), 
(\ref{expw}), (\ref{expu}) it follows then
that CP invariance constrains the matrix $m_D$ to satisfy :
\bea
{\rm arg}(m_D)_{ij}=\frac{1}{2}(\beta_i-\alpha_j) \label{arg}
\eea
Note that the $\alpha_i$ are fixed by Eq.~(\ref{ais}), up to discrete
ambiguities. Therefore CP invariance constrains the matrix $m_D$
to have only $n$ free phases $\beta_i$. Since $m_D$ is in an arbitrary
matrix, with
$n^2$ independent phases, it is clear that there are $n^2-n$ independent
CP restrictions. This number equals, of course, the number of independent
CP violating phases which appear in general in this model. In the WB
which
we are considering, these phases appear as $n(n-1)$ phases which cannot
be removed from $m_D$.
It is worth counting also the remaining physical parameters in this
WB in the case of three generations. 
The matrices $m_l$, $M_R$ are diagonal
and real and therefore they contain six real parameters. The matrix $m_D$
, apart from the six CP violating phases, has nine real parameters.
Thus one has a total of fifteen real
parameters and six CP violating phases.
We shall see that this counting of degrees of freedom agrees with the one
made in the physical basis, in terms of charged lepton and
neutrino masses together with the mixing angles and CP violating phases
entering in the leptonic mixing matrix.
\subsection{CP violating phases in the leptonic mixing matrix}
For definiteness, we shall consider the case of three generations
(three light neutrinos), where the full neutrino mass matrix, ${\cal M}$
in Eq.~(\ref{lm}), is
$ 6 \times 6$, and has the following form:
\bea
{\cal M}= \left(\begin{array}{cc}
0 & m \\
m^T & M \end{array}\right) \label{calm}
\eea
For notation simplicity, we have dropped the subscript in $m_D$
and $M_R$. The neutrino mass matrix is diagonalized by the
transformation:
\bea
V^T {\cal M}^* V = \cal D \label{dgm}
\eea
where ${\cal D} ={\rm diag.} (m_{\nu_1}, m_{\nu_2}, m_{\nu_3},
M_{\nu_1}, M_{\nu_2}, M_{\nu_3})$,
with $m_{\nu_i}$ and $M_{\nu_i}$ denoting the physical
masses of the light and heavy Majorana neutrinos, respectively. It is
convenient to write $V$ and $\cal D$ in the following form:
\bea
V&=&\left(\begin{array}{cc}
K & R \\
S & T \end{array}\right) ; \label{matv}\\
{\cal D}&=&\left(\begin{array}{cc}
d & 0 \\
0 & D \end{array}\right) . \label{matd}
\eea
From Eqs.~(\ref{calm}), (\ref{dgm}), (\ref{matv}), (\ref{matd}) one
obtains:
\bea
S^\dagger m^T K^* + K^\dagger m S^*
+ S^\dagger M S^* &=&d \label{12a} \\
S^\dagger m^T R^* + K^\dagger m T^*
+ S^\dagger M T^* &=&0 \label{12b} \\
T^\dagger m^T R^* + R^\dagger m T^*
+ T^\dagger M T^* &=&D  \label{12c}
\eea
From Eq.~(\ref{12b}) and taking into account 
that both S and R are of order 
$\frac{m}{M}$, one obtains, to an excellent approximation:
\bea
S^\dagger=-K^\dagger m M^{-1} \label{13}
\eea
From Eqs.~(\ref{12a}), (\ref{13}), it also follows
to an excellent approximation that:
\bea
-K^\dagger m \frac{1}{M} m^T K^* =d \label{14}
\eea
which is the usual seesaw formula. The neutrino weak-eigenstates are
related
to the mass eigenstates by:
\bea
{\nu^0_i}_L= V_{i \alpha} {\nu_{\alpha}}_L=(K, R)
\left(\begin{array}{c}
{\nu_i}_L  \\
{N_i}_L \end{array} \right) \quad \left(\begin{array}{c} i=1,2,3 \\
\alpha=1,2,...6 \end{array} \right)
\label{15}
\eea
and thus the leptonic charged current interactions are given by:\\
\bea
- \frac{g}{\sqrt{2}} \left( \overline{l_{iL}} \gamma_{\mu} K_{ij}
{\nu_j}_L +
\overline{l_{iL}} \gamma_{\mu} R_{ij} {N_j}_L \right) W^{\mu}+h.c.
\label{16}
\eea
From Eqs.~(\ref{15}), (\ref{16}) it follows that $K$ and $R$ give the
charged current
couplings of charged leptons to the light 
neutrinos $\nu_j$ and to the heavy
neutrinos $N_j$, respectively. In the exact decoupling limit, $R$
can be neglected and only $K$ is relevant. However, since we want to
study the connection between CP violation relevant to leptogenesis
and that detectable at low energies (e.g., in neutrino oscillations)
we have to keep both $K$ and $R$. Now, from the relation
${\cal M}^* V= V^* {\cal D}$ and taking into account the zero entry in
Eq.~(\ref{calm}), one derives the following exact relation:
\bea
R=m T^* D^{-1} \label{exa}
\eea
From Eq.~(\ref{12c}), and keeping in mind that we are working in a WB
where the
right-handed Majorana neutrino mass $M$ is diagonal, one concludes
that
$T= 1 $ up to corrections of order $\frac{m^2}{M^2}$. Therefore,
one
has, to an excellent approximation:
\bea
R=m D^{-1} \label{app}
\eea
From Eqs.~(\ref{14}), (\ref{app}), it is clear how the six physical
phases of the Dirac mass
matrix $m$, enter in the two blocks $K$, $R$ of the $ 3 \times 6$
leptonic mixing
matrix. It is useful to parametrize the Dirac neutrino mass matrix 
by \cite{Hashida:2000wh}:
\bea
m=U Y_{\triangle} \label{mui}
\eea
where $U$ is a unitary matrix and $Y_{\triangle}$ has a triangular
form:
\bea
Y_{\triangle}= \left(\begin{array}{ccc}
y_{1} & 0 & 0 \\
|y_{21}| \exp(i \phi_{21}) & y_{2} & 0 \\
|y_{31}| \exp(i \phi_{31}) & |y_{32}| \exp(i \phi_{32}) & y_{3}
\end{array}
\right) \label{tria}
\eea
where the $y_{i}$ are real. Since $U$ is unitary, it contains in
general
six phases. However, three of these phases can be rephased away through
the
transformation:
\bea
m \rightarrow P_{\xi} m  \label{mp}
\eea
where $P_{\xi}={\rm diag.}\left(\exp(i\xi_1),\exp(i\xi_2),\exp(i\xi_3)
\right)$. In a WB, this corresponds to a simultaneous phase
transformation
of the left-handed charged lepton fields and the 
left-handed neutrino fields. 
Furthermore, $Y_{\triangle}$ defined by Eq.~(\ref{tria}) can be
written as:
\bea
Y_{\triangle}= {P_{\beta}^\dagger}\ {\hat Y_{\triangle}}\ P_{\beta}
\label{pyp}
\eea
where $P_\beta =diag.(1, \exp(i\beta_1), \exp(i\beta_2))$ and
\bea
{\hat Y_{\triangle}}= \left(\begin{array}{ccc}
y_{1} & 0 & 0 \\
|y_{21}|  & y_{2} & 0 \\
|y_{31}|  & |y_{32}| \exp(i \sigma) & y_{3}
\end{array}
\right) \label{hat}
\eea
with $\sigma =  \phi_{32} -  \phi_{31} +  \phi_{21}$. It follows from 
Eqs.~(\ref{mui}), (\ref{pyp}) that the matrix  m can then be written as:
\bea
m={\hat U_{\rho}} P_{\alpha} {\hat Y_{\triangle}} P_{\beta} 
\label{upy}
\eea
where $P_\alpha =diag.(1, \exp(i\alpha_1), \exp(i\alpha_2))$ and
${\hat U}_{\rho}$ contains only one phase $\rho$ and can be written, for
example, through a ``standard" parametrization, as used in the case of the
CKM matrix \cite{Groom:2000in}. Therefore, in this WB, where $m_l$ and $M_R$ 
are diagonal and real,  the phases 
$\rho$, $\alpha_1$, $\alpha_2$,
$\sigma$, $\beta_{1}$, $\beta_{2}$ are the only physical phases 
and can be used to characterize CP violation in this model.
We shall see
in the sequel that leptogenesis is controlled by the phases
$\sigma$, $\beta_{1}$, $\beta_{2}$. 
Taking into consideration Eq.~(\ref{app}), which is valid to an
excellent approximation, one obtains:
\bea
R={\hat U_{\rho}} P_{\alpha} {\hat Y_{\triangle}} D^{-1} P_{\beta} 
\label{rup}
\eea
Note that the phases $\beta_{1}$ and  $\beta_{2}$ are of Majorana 
type in the  sense that they could be removed from $R$ through a
phase redefinition of the heavy Majorana neutrinos $N_{2}$, $N_{3}$.
Of course, since the $N_{i}$ are Majorana particles, this redefinition
would simply shift these phases
to the mass terms of $N_{2}$, $N_{3}$. 
At this point, the following comment
is in order. We are considering a simple extension of the SM, where there
are Dirac, as well as right-handed neutrino mass terms, but no
left-handed
Majorana mass terms. This is, of course, due to the absence of Higgs
triplets.
The fact that there are no left-handed Majorana neutrino 
mass terms leads to special
constraints. Indeed from Eqs.~(\ref{calm}), {(\ref{dgm}),
(\ref{matv}), (\ref{matd}),
one readily obtains the following
exact relation:
\bea
K^* d K^\dagger + R^* D R^\dagger =0. \label{krd}
\eea
In order to analyse the meaning of these constraints, let us for the
moment consider the most general case, i.e. the case in which a
left-handed Majorana mass term is also present.
The general $ 3 \times 6$ leptonic mixing matrix can then be exactly
parametrized by the first three rows of the
$ 6 \times 6$ unitary matrix $V$ which diagonalizes the full neutrino
mass
matrix ${\cal M}^*$, provided that $V$ is chosen in such a way that
a minimal number of phases appears in these first three rows. 
Such is the case with the following explicit parametrization for
$V$ \cite{Anselm:1985rw}:
\bea
V= {\hat V} P  \label{vip}
\eea
where $P={\rm diag.}
\left(1, \exp(i\sigma_1), \exp(i\sigma_2),..., \exp(i\sigma_5)
\right) $ and  $\hat V$ is given by:
\bea
{\hat V}=O_{56}I_6(\delta_{10}) O_{45} O_{46} I_5 (\delta_9)
I_6(\delta_8)...
O_{26} I_3(\delta_4)...I_6(\delta_1) O_{12}...O_{16} \label{vcp}
\eea
where $O_{ij}$ are orthogonal matrices mixing 
the ith and jth generation
and $I_j(\delta_k)$ are unitary diagonal matrices of the form:
\bea
I_j(\delta_k)=\left(\begin{array}{cccccccc} \\
1&&&&&& \\
&.&&&&& \\
&&1&&&& \\
&&&e^{i \delta_k}&&&\\
&&&&1&& \\
&&&&&.& \\
&&&&&&1 \\
\end{array} \right) \leftarrow j
\label{ij}
\eea
It can be readily verified that the first three rows of 
$\hat V$, contain seven phases.
This parametrization is particularly useful, for
instance, in models with vectorial quarks \cite{Branco:1986my}.
Together with the five phases contained in $P$, one
has, in the general case, twelve phases characterizing the $ 3 \times 6$
leptonic mixing matrix.
The previous counting shows that the constraint of Eq.~(\ref{krd})
leads to the decrease
of the number of independent phases from twelve to six. This is to be
expected,
since in the general case, the left-hand side
of Eq.~(\ref{krd}) equals the
left-handed Majorana matrices mass $m_{\nu}$, a $3 \times 3$ complex
symmetric matrix, which in general contains six
phases. Therefore, putting $m_{\nu}=0$
implies the loss of six independent phases. Of course,
one can still use the parametrization defined by 
Eqs.~(\ref{vip}), (\ref{vcp}), (\ref{ij}) 
but the angles and phases
introduced are not independent parameters,
since they are constrained by Eq.~(\ref{krd}).
\section{Weak-basis invariants and CP violation}
\setcounter{equation}{0}
In section 2.1, we have analysed CP violation in a particular WB. We
shall derive now WB invariants which have to vanish 
in order for CP invariance to hold.
The non-vanishing of any of these 
invariants signals CP violation \cite{Branco:1986gr}.
We are specially interested in WB invariants  sensitive to the
CP violating phases which appear in leptogenesis.
From Eqs.~(\ref{cpm}), (\ref{cpM}), one obtains:
\bea
W^{\dagger}h W&=& h^* \nn  \\
W^{\dagger}H W&=& H^* \label{wh}
\eea
where $h=m^{\dagger}m$, $H=M^{\dagger}M$.
It can be then readily derived, from Eqs.~(\ref{cpM}),
(\ref{wh}), that CP invariance requires:
\bea
I_1 \equiv {\rm Im Tr}[h H M^* h^* M]=0 \label{i1}
\eea
Since $I_1$ is a WB invariant, it may be evaluated in any convenient WB.
In the WB where the right-handed neutrino mass $M$ is diagonal, one
obtains:
\bea
I_1 &=& M_1 M_2 ({M_2}^2-{M_1}^2) {\rm Im}({h_{12}}^2)+M_1 M_3
({M_3}^2-{M_1}^2){\rm Im}( {h_{13}}^2) + \nonumber \\
&+& M_2 M_3 ({M_3}^2-{M_2}^2) {\rm Im}( {h_{23}}^2)=0
\label{i1l}
\eea
The appearance of the
quadratic term ${h_{ij}}^2$ was to be expected since it reflects the well
known fact that phases of $\frac{\pi}{2}$ in $h_{ij}$ do not imply CP
violation.
The interest of $I_1$ stems from the fact that, as will be seen
in section 4, the strength of leptogenesis crucially depends on
${\rm Im}({h_{ij}}^2)$ $( j \ne i) $.
Similarly, one can readily show that CP invariance also implies
the vanishing of the WB invariants $I_2$, $I_3$:
\begin{eqnarray}
I_2 &\equiv& {\rm Im Tr}[h H^2 M^* h^* M] = \nonumber \\
&=& M_1 M_2 ({M_2}^4-{M_1}^4) {\rm Im}({h_{12}}^2)+M_1 M_3
({M_3}^4-{M_1}^4){\rm Im}( {h_{13}}^2) + \nonumber \\
&+& M_2 M_3 ({M_3}^4-{M_2}^4) {\rm Im}( {h_{23}}^2)=0
\ \
\label {i2l}\\
I_3 &\equiv& {\rm Im Tr}[h H^2 M^* h^* M H] = \nonumber  \\
&=&{M_1}^3 {M_2}^3 ({M_2}^2-{M_1}^2) {\rm Im}({h_{12}}^2)
+{M_1}^3 {M_3}^3 ({M_3}^2-{M_1}^2) {\rm Im}( {h_{13}}^2) + \nonumber \\
&+& {M_2}^3 {M_3}^3 ({M_3}^2-{M_2}^2) {\rm Im}( {h_{23}}^2)=0 \label{i3l}
\end{eqnarray}
where again we have given the explicit expressions for $I_2$, $I_3$, in
the basis where M is real, diagonal. Note that the three
Eqs.~(\ref{i1l}), (\ref{i2l}), (\ref{i3l})
constitute a set of linear equations in terms of the variables
${\rm Im}({h_{ij}}^2)$, where
the coefficients are functions of the right-handed neutrino masses $M_i$.
The determinant of the coefficients of this set of equations can be
readily evaluated and one obtains:
\bea
Det.={M_1}^2 {M_2}^2 {M_3}^2 
{\Delta^2}_{21}{\Delta^2}_{31}{\Delta^2}_{32}
\label{dmd}
\eea
where ${\Delta}_{ij}=({M_i}^2-{M_j}^2)$.
From Eq.~(\ref{dmd}), it follows that if none of the $M_i$ vanish and
furthermore there is no degeneracy, the vanishing of $I_1, I_2, I_3$
implies the vanishing of ${\rm Im} ({h_{12}}^2)$,
${\rm Im} ({h_{13}}^2)$, ${\rm Im} ({h_{23}}^2)$.
Since there are six independent CP violating phases, one may wonder
whether
one can construct other three independent WB invariants, apart from
$I_i$,
which would describe CP violation in the leptonic sector. This is indeed
possible, a simple choice are the WB invariants ${\bar I}_i (i=1,2,3)$,
obtained
from $I_i$, through the substitution of $h$ by ${\bar h}=m^{\dagger} h_l
m$,
where $h_l=m_l {m_l}^{\dagger}$. For example one has:
\bea
{\bar I_1}={\rm Im Tr}(m^{\dagger}h_l m H M^* m^T {h_l}^* m^* M)
\label{ibar}
\eea
and similarly for $\bar{I_2}, \bar{I_3}$. As it was the case for $I_i$,
CP invariance requires that $\bar{I_i}=0$. And likewise, the vanishing of
$\bar{I_i}$ implies (barring either vanishing or degenerate $M_i$) that
${\rm Im} ({\bar{h_{ij}}}^2)$ $(i \ne j)$ vanish.
\section{ CP violating phases relevant for leptogenesis}
\setcounter{equation}{0}
In this section, we identify the CP violating phases relevant for
leptogenesis, obtained through the out of equilibrium decay of heavy
Majorana neutrinos.
In the previous section, we have seen that
there are six independent CP violation sources in the minimal
seesaw model by constructing WB invariants.
Next we show which of these six independent CP violating
sources contribute to lepton number asymmetry.
\subsection{ Lepton number asymmetry}
Leptogenesis gives rise to the BAU through the out of
the equilibrium decay of heavy Majorana neutrinos in
the symmetric phase, 
with the generation of $L \neq 0$ while $B = 0$ is
still maintained. Later on, 
sphaleron processes \cite{Kuzmin:1985mm}
restore $B+L =0$ in the universe leaving $B-L$ invariant, 
thus creating
a non vanishing $B$. Because the 
lepton number is generated at a very
high temperature in this scenario, the lepton number
asymmetry is generated in the symmetric phase, i.e., $v=0$, where
$v$ is the Higgs vacuum expectation value.
In the previous sections,
we started with the Lagrangian in the broken phase 
and identified the CP
violating quantities. In order
to see the connection between the lepton number asymmetry
generated in the symmetric phase and
the CP violating quantities defined in the broken phase, 
i.e., the phases
in the $ 3 \times 6$ mixing matrix and the WB invariant
defined in terms of the Dirac mass matrix m
and the Majorana mass matrix M,
we use the Lagrangian in the broken
phase for the computation of the asymmetry in the symmetric phase.
This is possible by simply taking the symmetric limit, 
i.e., $v \rightarrow 0$
in the broken phase computation.
By following this procedure, not only do we recover 
the known result in the symmetric
phase, but we also clarify the relation between
CP violation generating lepton asymmetry 
at a high energy and CP violation
in the broken phase.
To be definite, we compute the decay of a heavy Majorana
neutrino $N_j$ into charged leptons ${l_i}^{\pm} $. The light
flavour indices $i=1,2$ and $3$ correspond
to $e$, $\mu$ and $\tau$ respectively. We can
define the lepton family number asymmetry as $\Delta {A^j}_i={N^j}_i-
{{\overline{N}}^j}_i$ and the lepton number asymmetry is obtained
by summing over the three flavours:
\bea
\Delta {A^j} =\sum_i \Delta {N^j}_i. \label{daj}
\eea
In the symmetric phase, the heavy Majorana
neutrinos decay into charged leptons and
charged Higgs. In the broken phase, the charged 
Higgs boson is absorbed into the
$W$ boson and the decay of heavy Majorana neutrinos into charged
leptons and charged Higgs
has no physical significance. In the broken phase,
the decay into $W$ bosons and charged leptons is the process
generating lepton number asymmetry. Of all possible polarizations
of the $W$ boson as final states, the longitudinal gauge boson gives the
dominant contribution.
Because the asymmetry comes from the interference between the
tree amplitude and the
absorptive part of the one-loop amplitudes, we can write the
amplitude as follows:
\bea
&&M(N_{j} \rightarrow {l_i}^{+} W^-)
=\langle {{l_i}^{+}} W^-|T|N_{j} \rangle \nn \\
&+&i \pi \sum \langle {l_i}^{+} W^-|T|{l_n}^{-} W^+ \rangle
\langle {l_n}^{-} W^+|T|N_{j} \rangle
\delta(E_n+E_W-M_{j}) \nn \\
&+&i \pi \sum \langle {l_i}^{+} W^-|T|\nu_{n} Z \rangle
\langle \nu_{n} Z|T|N_{j} \rangle
\delta(E_n+E_Z-M_{j})
\nn \\
&+&i \pi \sum \langle {l_i}^{+} W^-|T|\nu_{n} H \rangle
\langle \nu_{n} H|T|N_{j} \rangle
\delta(E_n+E_H-M_{j}), \label{nlw}
\eea
where $\langle T \rangle $ denotes the tree amplitudes.
We compute the absorptive part by summing over possible
two body on-shell states such as
$l^- W^+$, $\nu Z$
and $\nu H$,
so that $\sum $ stands for
the sum over flavour indices $n$, three momentum $q$, polarization,
and spin. For example, for a $W$ and a charged lepton state,
$\sum$ means $\int \sum_{pol,spin,n}
\frac{d^3 q}{2E_n 2E_W (2 \pi)^3}$.
The following analysis
shows that of all the polarizations, the longitudinal
$W$ and $Z$ boson dominate in the sum. In the rest frame of the
heavy Majorana neutrino,
the gauge bosons which appear in the matrix elements
carry half of the energy and the momentum of
the decaying heavy Majorana neutrino, i.e.,
$E_W \simeq P_W \simeq \frac{M}{2}$. Therefore,
in leading order of
the $\frac{v}{M}$ expansion, the polarization of
the longitudinal $W$ $(W_L)$ is written as:
\bea
{\epsilon_L}^{\mu}&&\simeq \frac{{P_W}^{\mu}}{M_W}. \label{elu}
\eea
By substituting it into charged current interaction, we obtain for the
tree level matrix element of $N_j$ to ${l_i}^-$ ${W_L}^+$ :
\bea
M^{\rm tree}(N_j \rightarrow {W_L}^+ {l_i}^{-})&=&
-R_{i j} \frac{g}{\sqrt{2}} (\epsilon_L)^{\mu *}
{\bar{u_i}} \gamma_{\mu} L U_{N}, \nn \\
&=& -R_{ij} M_j \frac{g}{\sqrt{2}M_W} {\bar{u_i}} R U_{N},
\label{tree}
\eea
where we use $P_W=P_j-P_i$ and neglect the lepton mass.
We note that only the interaction to the longitudinal polarization
remains in the small $\frac{v}{M}$ limit:
\bea
R_{ij} M_j \frac{g}{\sqrt{2} M_W} \rightarrow
\frac{\sqrt{2}{m }_{ij}}{v}=
y_{Dij}, \label{rmg}
\eea
with $y_{Dij}$ the coefficients of the neutrino Yukawa couplings.
In the following discussion, we only keep the longitudinal polarization
for $W$ and $Z$. Including the Higgs boson interaction, 
the relevant part of
the Lagrangian needed for the computation of the
matrix elements in Eq.~(\ref{nlw}), is given by:
\bea
{\cal L}&=& -\frac{g}{\sqrt{2}}{(R^{\dagger})}_{ji}
(\bar{N_{j}}\gamma_{\mu}{l_{i}}_L )
W^{\mu +}
-\frac{g}{2 \cos \theta_W }{(R^{\dagger} K)}_{ji}
(\bar{N_{j}}\gamma_{\mu}{\nu_{i}}_L )
Z^{\mu} \nn \\
&& -\frac{g}{2 M_W} M_j {(R^{\dagger} K)}_{ji}
(\bar{N_{j}}{\nu_{i}}_L )
H
\nn \\
&& -\frac{g}{\sqrt{2}} R_{ij}
(\overline{{l_i}_L}\gamma_{\mu} N_{j} )W^{\mu -}
-\frac{g}{2 \cos \theta_W }
{(K^{\dagger} R)}_{ij}(\overline{{\nu_i}_L}\gamma_{\mu} N_{j})
Z^{\mu} \nn \\
&&-\frac{g}{2 M_W }{(K^{\dagger} R)}_{ij} M_j
(\overline{{\nu_i}_L} N_{j} ) H. \label{lag}
\eea
By neglecting the masses of the Higgs, $W$, $Z$, 
charged leptons, and light neutrinos
compared to the heavy Majorana neutrinos masses,
we obtain the following result:
\bea
&& M(N_j \rightarrow {l_{i}}^+ {W_L}^-) \nn \\
&=&(\bar{u_i} L U_N) \frac{g M_j}{\sqrt{2} M_W}
\Bigl[{(R^{\dagger})}_{ji}-i {(R^{\dagger})}_{kn}{(R^{\dagger})}_{ki}
R_{nj}
\left(\frac{g M_k}{\sqrt{2} M_W}\right)^2 \frac{1}{16 \pi}
\left(I(x_k)+ \frac{\sqrt{x_k}}{2(1-x_k)} \right) \nn \\
&& -i {(R^{\dagger}K)}_{kn}{(R^{\dagger})}_{ki} (K^{\dagger}R)_{nj}
\left(\frac{g M_k}{\sqrt{2} M_W}\right)^2 \frac{1}{16 \pi}
\left(\frac{\sqrt{x_k}}{2(1-x_k)} \right) \Bigr]+...,
\nn \\
&=&(\bar{u_i} L U_N) \frac{g M_j}{\sqrt{2} M_W}
\Bigl[{(R^{\dagger})}_{ji}-i {(R^{\dagger})}_{kn}{(R^{\dagger})}_{ki}
R_{nj}
\left(\frac{g M_k}{\sqrt{2} M_W}\right)^2 \frac{1}{16 \pi}
\left(I(x_k)+ \frac{\sqrt{x_k}}{(1-x_k)} \right) \Biggr] \nn \\
&+&
i g^3 \left( (R^{\dagger})_{ji}... + (R^{\dagger} R)_{jk}
(R^{\dagger})_{ki}...
\right), \label{exp}
\eea
where $x_k=\frac{{M_k}^2}{{M_j}^2}$ and
$ I(x_k)=\sqrt{x_k} \left(1+(1+x_k) \log(\frac{x_k}{1+x_k}) \right)$.
We have used $K^{\dagger}K =1$ which holds up to $O(\frac{v^2}{M^2})$.
$M_k$ denotes the mass of the $k$ th heavy Majorana neutrino
which contributes to s channel and t channel scattering amplitudes
denoted by $ \langle l W|T| l W \rangle $,
$ \langle l Z|T| l W \rangle $ and $ \langle l H|T| l W \rangle $.
In Eq.~(\ref{exp}), the first term comes from the tree level
amplitude the second term comes from the
absorptive part of $W$ and charged lepton
and the third term comes from sum of the absorptive part of
Z and neutrino, and Higgs and neutrino. The terms indicated by $...$
do not contribute to the total lepton asymmetry, to leading order,
because one has 
${\rm Im} R_{ij} (R^{\dagger})_{ji}=0$ and
${\rm Im} (R^{\dagger} R)_{jk} (R^{\dagger} R)_{kj}=0$.
A similar formula can be obtained for the $l^- W^+$ final state:
\bea
&&M(N_j \rightarrow {l_i}^{-} {W_L}^+)= \nn \\
&& - (\bar{u_i} R U_N)\frac{g M_j}{\sqrt{2} M_W}
\left[R_{ij}-i R_{nk}R_{ik}
(R^{\dagger})_{jn}
\left(\frac{g M_k}{\sqrt{2} M_W}\right)^2 \frac{1}{16 \pi}
\left(I(x_k)+ \frac{\sqrt{x_k}}{1-x_k} \right) \right]. \nn \\
\label{urn}
\eea
The lepton number asymmetry from j th heavy Majorana particle
is then given by:
\bea
A^j&=&\frac{\sum_i \Delta {A^j}_i}{\sum_i \left({N^j}_i +
\overline{N^j}_i \right)} \nn \\
&=& \frac{g^2}{{M_W}^2} \sum_{k \ne j} \left[ (M_k)^2
{\rm Im} \left((R^\dagger R)_{jk} (R^\dagger R)_{jk} \right)
\frac{1}{16 \pi} \left(I(x_k)+ \frac{\sqrt{x_k}}{1-x_k} \right)
\right]
\frac{1}{(R^\dagger R)_{jj}} \nn \\
&=& \frac{g^2}{{M_W}^2} \sum_{k \ne j} \left[
{\rm Im} \left((m^\dagger m)_{jk} (m^\dagger m)_{jk} \right)
\frac{1}{16 \pi} \left(I(x_k)+ \frac{\sqrt{x_k}}{1-x_k} \right)
\right]
\frac{1}{(m^\dagger m)_{jj}}, \nn \\
&=& \sum_{k \ne j} \left[
{\rm Im} \left(({y_D}^\dagger y_D)_{jk} ({y_D}^\dagger y_D)_{jk}
\right)
\frac{1}{8 \pi} \left(I(x_k)+ \frac{\sqrt{x_k}}{1-x_k} \right)
\right]
\frac{1}{({y_D}^\dagger y_D)_{jj}}, \label{rmy}
\eea
where in the third equality, we used the approximate formulae:
$ R_{ij} M_j = m_{ij}$ and in the final expression, 
we substituted
$m_{ij}=y_{D ij} \frac{v}{\sqrt{2}}$.
This expression for $A^j$ agrees with the one computed in
the symmetric phase \cite{sym}.  
The expression in terms of the 
mixing matrix $R$ was first derived, in a different way,
in \cite{Endoh:2001hc} without using the unitary
gauge as we have done in this paper.
To summarize our results, starting with
the Lagrangian in the broken phase, 
we have computed the lepton number asymmetry and 
in the limit $v \rightarrow 0$, 
we have exactly recovered the
lepton number asymmetry computed in the symmetric phase. 
The following argument shows that 
this result was to be expected.
In the symmetric phase, the Nambu-Goldstone
boson becomes physical and the gauge bosons become massless. The heavy
Majorana neutrino decays through the Yukawa coupling to the would-be
Nambu-Goldstone boson. In the broken phase,
the gauge bosons become massive and the Nambu-Goldstone boson is absorbed
as their longitudinal component. 
The heavy Majorana neutrino can decay
into both transverse and longitudinal gauge bosons. The interaction
to the transverse polarization is suppressed by the mixing $R$ and can be
neglected for small  $\frac{v}{M}$, while the
coupling to the longitudinal polarization remains finite in the limit
of $\frac{v}{M} \rightarrow 0$ and  this
is given by the original Yukawa coupling (See Eq.~(\ref{rmg})).
This is the essential reason why we
obtain the same result as the one found in the symmetric phase
although we started with the broken phase.
{\subsection {CP violation in lepton number asymmetry}}
From Eq.~(\ref{rmy}) together with the parametrization chosen in
Eq.~(\ref{upy}) it can be seen that the
lepton number asymmetry is only sensitive to the CP violating
phases appearing in $Y_{\Delta}$, since
\bea
{\rm Im} \{ (m^{\dagger} m)_{jk} \}^2&=&
{\rm Im} \{ ({Y_{\Delta}}^{\dagger} Y_{\Delta})_{jk} \}^2, \quad (j \ne
k).
\eea
Explicitly, one has:
\bea
(m^{\dagger} m)_{12}&=&|y_{21}| y_2 \exp (i \beta_{1})+
|y_{31} y_{32}| \exp \left(i (\beta_{1} + \sigma) \right) \nn \\
(m^{\dagger} m)_{23}&=&|y_{32}| y_3 \exp (-i (\beta_{2} -
\beta_{1} -\sigma))
\nn \\
(m^{\dagger} m)_{31}&=&|y_{31}| y_3 \exp(-i \beta_{2}).
\eea
Thus the three phases $\sigma$, $ \beta_{1}$ and $\beta_{2}$
appearing in $Y_{\Delta}$ generate the lepton number
asymmetry \cite{Endoh:2001hc}.
From our analysis in section 3, it is clear that the lepton asymmetry
can be related to WB invariants.
We have seen that there are
six independent WB invariants which signal CP violation in the minimal
seesaw model. The combinations which generate lepton
number asymmetry can be written in terms of $h = m^\dagger m$:
\bea
{\rm Im} \{(m^{\dagger}m)_{jk}\}^2= {\rm Im} (h_{jk})^2.
\eea
where $j\ne k$.
By going to the WB, in which both charged lepton and heavy
Majorana mass matrices are real diagonal, we have shown that
the three WB invariants $I_1 \sim I_3$ are functions of
${\rm Im}({h_{12}}^2)$, ${\rm Im}({h_{23}}^2)$ 
and ${\rm Im}({h_{31}}^2)$.
Therefore $A^j$ can be written in terms of WB invariants $I_1$, $I_2$ and
$I_3$
and is independent of the other three WB invariants ${\bar I}_i \ 
(i=1,2,3)$.
\section{Relating CP violation in leptogenesis with CP violation at low
energies}
\setcounter{equation}{0}
\subsection{Model independent analysis}
One of the most fascinating questions one may ask is whether there is
some connection between CP violation responsible for the generation of
BAU through leptogenesis and the one measurable at low energies.
More specifically, assuming that BAU is indeed achieved through
leptogenesis, one may ask what can
one infer about the size of CP violation at low energies, from the size
and
sign of the observed BAU. In order to address the above question, one
should
keep in mind that in the WB where both the right-handed neutrino mass
matrix
M and the charged lepton mass matrix $m_l$ are diagonal, real, all
information
about the leptonic mixing and CP violation is contained in the Dirac
neutrino
mass matrix m. We have seen that in the minimal seesaw model we are
considering,
the matrix $m$ contains six phases. For definiteness, let us consider the
parametrization of $m$ given by Eq.~(\ref{upy}), where the six phases are
$\rho$, $\alpha_1$, $\alpha_2$, $\sigma$, $\beta_{1}$, $\beta_{2}$.
In general and without
further assumptions about the structure of the leptonic mass matrices,
these
six phases are independent from each other. Furthermore, we have seen in
section 4 that leptogenesis is controlled by the phases
$\sigma$, $\beta_{1}$ and  $\beta_{2}$. In order to see what are the
phases
relevant to CP violation at low energies,
we have to consider the effective left-handed
neutrino mass matrix, given by:
\bea
m_{ef}=-m \frac{1}{D} m^T =-{\hat U_{\rho}} P_{\alpha} 
{\hat Y_{\triangle}} {P_{\beta}}^2
\frac{1}{D}
{\hat Y_{\triangle}}^T {P_{\alpha}} {\hat U_{\rho}}^T  \label{mef}
\eea
The strength of CP violation at low energies, observable 
for example through neutrino
oscillations, can be obtained from the following low-energy WB invariant:
\bea
Tr[h_{ef}, h_l]^3=6i \Delta_{21} \Delta_{32} \Delta_{31}
{\rm Im} \{ (h_{ef})_{12}(h_{ef})_{23}(h_{ef})_{31} \} \label{trc}
\eea
where $h_{ef}=m_{ef}{m_{ef}}^{\dagger},\  h_l=m_l {m_l}^{\dagger}$ and
$\Delta_{21}=({m_{\mu}}^2-{m_e}^2)$ with analogous expressions for
$\Delta_{31}$, $\Delta_{32}$.
From Eqs.~(\ref{mef}), (\ref{trc}), it is clear that all
the six phases contained in $m$ affect the strength of CP violation at
low energies. In the decoupling limit, leptonic mixing is characterized,
of course, by a $3 \times 3$ unitary matrix, containing one CP violating
Dirac phase $\delta$ and two Majorana phases, which have an interesting 
geometrical interpretation in terms of unitarity triangles
\cite{Aguilar-Saavedra:2000vr}. CP violation in neutrino oscillations
is only affected by the phase $\delta$. The important point is that the
phase
$\delta$ is, in general, a function of all the six phases $\rho$,
$\alpha_1$, $\alpha_2$,
$\sigma$, $\beta_{1}$, $\beta_{2}$ as can be seen 
from Eqs.~(\ref{14}), (\ref{upy}). 
Since leptogenesis only depends  on $\sigma $, $\beta_{1}$ and 
$\beta_{2}$,
it is clear that, in general, one cannot directly relate the size
of CP violation responsible for leptogenesis with the strength 
of CP violation at low energies. More specifically, one cannot
derive the size of the phase $\delta $ from the knowledge of 
the amount of CP violation required to generate the observed 
baryon asymmetry, through leptogenesis. A related important 
question is the following. 
Throughout the paper, we have assumed completely 
generic leptonic mass matrices. Let us now consider a scheme where 
the only non-vanishing phases are those responsible for
leptogenesis, namely $\sigma$, $\beta_{1}$ and $\beta_{2}$. Can one
in such a scheme, generate CP violation at low energies?
In order to answer this question, one has to compute the WB
invariant  $Tr[h_{ef}, h_l]^3$ given in Eq.~(\ref{trc}), and,
in particular, $ {\rm Im}Q =  {\rm Im}
\{ (h_{ef})_{12}(h_{ef})_{23}(h_{ef})_{31} \}$. 
Through a tedious but
straightforward computation one can verify that ${\rm Im} Q$
does not vanish in general. This is true even if only one 
of the phases $\sigma$, $\beta_{1}$ or $\beta_{2}$ is non-vanishing.
One concludes then that in a model where the leptonic
mass matrices are constrained (e. g. by flavour symmetries) so that 
only one of the phases (for example $\sigma$ ) is non-vanishing,
one can establish a direct connection between the size of the observed
BAU and the strength of CP violation at low energies, observable
for example in neutrino oscillations. In section 5.2 we shall
describe a framework where the only 
non-vanishing phases are those responsible for leptogenesis.
\subsection{A special GUT inspired scenario}
In the previous subsection, our analysis, done in the 
framework of the minimal seesaw, has been model 
independent and we have
discussed, with all generality, the phases which can be seen
by leptogenesis, as well as those appearing at low energy
CP violating processes. Let us now assume that the
Hermitian matrices $m_l m_l^{\dagger}$, $m m^{\dagger}$
are diagonalized by the same left-handed unitary transformation,
a situation which occurs in some
GUTs. In this case, in the WB where $m_l$ and the
rigt-handed Majorana neutrino matrix M are both real 
and diagonal, the neutrino Dirac mass matrix has the form:
\bea
m=dU_{R}  \label{dur}
\eea
where d is diagonal and $U_R$ is a generic unitary matrix.
It can be easily seen that in this case there are only three 
independent phases. In fact the phases of $U_R$ 
corresponding to $P_{\xi}$ in Eq.~(\ref{mp}) can be eliminated
in the same way, due to  commutativity with the
diagonal matrix d. Two of the remaining three phases in 
$U_R$ can be factored out and we may write:
\bea
m=d{\hat U}(\delta ) P_{\alpha}, \ \ \ 
P_{\alpha} = diag(1, \exp (i \alpha_{1}), \exp (i \alpha_{2}))
\label{dup}
\eea
We conclude from Eq.~(\ref{app}) that in the limit $T=1$
the phases $\alpha_{1}$ and $\alpha_{2}$ are seen,
in the heavy neutrino sector, as Majorana type phases in 
analogy to the phases $\beta_{1}$ and $\beta_{2}$
in Eq.~(\ref{rup}). On the other hand ${\hat U}(\delta)$
can be parametrized as the CKM matrix with only one phase 
$\delta$. As a result, the matrix m will contain only three
phases, namely the phases $\delta$, $\alpha_{1}$ and
$\alpha_{2}$. In section 4 we pointed out that leptogenesis 
is only sensitive to the phases appearing in $m^{\dagger} m$
and so it is clear that these three phases are exactly those
appearing in leptogenesis. Furthermore it can be 
readily verified that these phases do lead to CP violation
at low energies. This is due to the fact that from 
Eq.~(\ref{dur}), one obtains for the effective lefthanded 
neutrino mass matrix:
\bea
m_{ef}= - d U_{R} \frac{1}{D} U_{R}^T d
\label{dud}
\eea
where D is the matrix M in the WB where it is diagonal.
The crucial point is that the phases in $U_R$ do not cancel
out in Eq.~(\ref{dud}). It is well known 
\cite{Altarelli:2000dg}, \cite{Akhmedov:2001yt} that  
in such a framework the large neutrino mixing required to
account for the atmospheric neutrino anomaly can be 
generated even for small mixing in ${\hat U}(\delta)$,
with hierarchical entries in D. Specific examples can be
constructed and even in the case of a single nonvanishing
phase, one still generates CP violation at low energies
- this is the case for only $\delta \neq 0$ or in 
alternative, for instance, only $\alpha_{2} \neq 0$.
These features are particularly striking since, 
as was pointed out above, these phases appear in the
heavy sector with a different nature (Dirac versus 
Majorana type).

There is an alternative WB for this scenario where 
m and $m_l$ are diagonalized simultaneously and M becomes:
\bea
M=U_R^* D U_R^\dagger  \label{alt}
\eea
As a result, one might view all CP violating effects
as related to the phases $\delta$, $\alpha_{1}$
and $\alpha_{2}$ generated at high energies. It
is common in the literature to assume hierarchical
entries in D. One might wonder whether this choice
plays a crucial r\^ ole in the connection between CP 
violation at high energies and CP violation at low
energies. In order to answer this question it is 
worthwhile computing 
${\rm Im} \{ (h_{ef})_{12}(h_{ef})_{23}(h_{ef})_{31} \}$
in the case of exact degeneracy of M. In the WB where
$m=d$ and $m_l$ is diagonal we have:
\bea
M=U_R^* D U_R^\dagger = \mu U_R^* U_R^{\dagger} 
=\mu Z_{0}  \label{muz}
\eea
with  $\mu$ the common degenerate mass and $Z_{0}$
a symmetric unitary matrix. It was shown in
Ref. \cite{Branco:1999bw} that one may have 
a nontrivial M, even in the limit of exact degeneracy,
which can be parametrized by:
\bea
M=\mu Z_{0} = \mu \left( \begin{array}{ccc}
1 & 0 & 0 \\
0 & c\phi & s\phi\\
0 & s\phi & -c\phi
\end{array}
\right)
\left( \begin{array}{ccc}
c\theta & s\theta & 0\\
s\theta & -c\theta & 0\\
0 & 0 & e^{i\alpha}
\end{array}
\right)
\left( \begin{array}{ccc}
1 & 0 & 0\\
0 & c\phi & s\phi\\
0 & s\phi & -c\phi
\end{array}
\right) \label{mmm}
\eea
where c, s stand for cosine and sine. 
In the case of full degeneracy, there are only two
independent angles $\phi$ and $\theta$ 
and one CP violating phase, $\alpha$. It can be easily
checked that:
\bea
& {\rm Im} \{ (h_{ef})_{12} (h_{ef})_{23} (h_{ef})_{31} \} =
d_1^{2} d_2^{2} d_3^{2} (d_1^{2} - d_2^{2}) 
(d_2^{2} - d_3^{2}) (d_1^{2} - d_3^{2})
\frac{1}{\mu ^6} s\alpha c\theta s^{2} \theta
s^{2} \phi c^{2} \phi = \nn  \\
& = d_1^{2} d_2^{2} d_3^{2} (d_1^{2} - d_2^{2})
(d_2^{2} - d_3^{2}) (d_1^{2} - d_3^{2})
\frac{1}{\mu^6} \left( \frac{1}{\mu^4} {\rm Im}
(M_{11} M_{12}^* M_{21}^* M_{22}) \right)
\label{imq}
\eea
where the $d_i$'s are the diagonal elements of $m=d$
and the imaginary part of a quartet of the unitary 
matrix M appears. This result means that even in this
very special limit of exact degeneracy in M,
CP violation at high energies 
can lead to CP violation at low energies. 
Note that in a framework where the 
CP violating phases appear
only in M, CP is only softly broken by the Lagrangian.
\subsection{A model with spontaneous CP violation at 
a high energy scale}
So far, in our analysis, we have not dealt with the
origin of the CP violating phases. These phases 
can appear as explicit CP violating terms in the 
Lagrangian, either through complex Yukawa couplings
and/or complex entries in the right-handed neutrino
Majorana mass M, or, in alternative, as the result of 
spontaneous CP violation. In this subsection,
we describe a minimal extension of the SM, 
where CP is a good symmetry of the Lagrangian,
only broken by the vacuum. We show that the model
has the interesting feature that the breaking
of CP originates at a high energy scale,
through a single phase of a complex scalar 
field which is sufficient to generate CP violation 
necessary for leptogenesis and at the same time 
generate CP violation at low energies.
Let us consider an extension of the SM where
one adds one right-handed neutrino per generation,
as well as a complex scalar S, invariant under
$SU(3)_c \times SU(2)_L \times U(1)$. The most 
general Yukawa and mass terms invariant under
$SU(3)_c \times SU(2)_L \times U(1)$ can be written:
\bea
{\cal L}_Y & = & \left( \overline{{\nu}^0},
\overline{e^0} \right)_L Y_l \left[ \begin{array}
{c} {\phi}^+ \\ {\phi}^0 \end{array} \right] e^0_R +
\left( \overline{{\nu}^0}, \overline{e^0}
\right)_L Y_{\nu} \left[ \begin{array}{c}
{\phi}^{0*} \\ -{\phi}^- 
\end{array} \right] {\nu}^0_R + \nn \\
& + & \frac{1}{2} \nu_R^{0T} {\overline M} \nu_R^0 +
\frac{1}{2} Y_S \nu_R^{0T} C \nu_R^0 S + \frac{1}{2}
Y^\prime_S \nu_R^{0T} C \nu_R^0 S^* + h.c.  
\label{lon}
\eea
Since we impose CP invariance on the Lagrangian,
all the Yukawa couplings $Y_l$, $Y_\nu$,
$Y_S$, ${Y^\prime}_S$, as well as the mass term 
$\overline M$ are real. Due to the presence 
in the Higgs potential of terms like
$S^2$, ${S^*}^2$, $S^4$, ${S^*}^4$, it can 
be readily seen that there is a region of the
parameters of the Higgs potential, where
the minimum is at:
\bea
\langle \phi^0 \rangle = v ;\ \ \  
\langle S \rangle = V e^{i \theta}
\label{fis}
\eea
The following mass terms are thus generated:
\bea
m_l = v Y_l; \ \ \ m = v Y_{\nu};\ \ \  
M = {\overline M} + Y_S V e^{i \theta} +
{Y^\prime}_S V e^{-i \theta} \label{xxx}
\eea
Since the Yukawa terms $Y_S$, ${Y^\prime}_S$
are real, but arbitrary, M is a general 
complex symmetric matrix, while $m_l$,
$m_D$ are real. It is clear that one can 
go to a WB where $m_l$, M are real, diagonal 
while the neutrino Dirac mass matrix is of the form:
\bea
m={O_L}^T d U_R   \label{olt}
\eea
with $O_L$ an orthogonal matrix, while $U_R$
is a generic unitary matrix. The CP violating 
phases contained in $U_R$ arise from the 
diagonalization of M and therefore these phases have
their origin in the vacuum phase $\theta$ of 
Eq.~(\ref{fis}). Of course, $U_R$ can be written as 
${P^\prime}_R {\hat U_{R}} (\phi) P_R$, with
${P^\prime}_R$, $P_R$ diagonal unitary matrices. 
However, it should be noted that, due to the presence of 
the non-trivial orthogonal matrix $O_L$, 
it is not possible to
eliminate the three phases contained in ${P^\prime}_R$.

Although CP violation arises in this model
from the single vacuum phase $\theta$,
it is clear that in general one has CP violation
necessary to obtain leptogenesis (with the
relevant phases being the three phases 
appearing in $m^\dagger m = {U_R}^\dagger d^2 U_R$)
as well as CP violation at low energies.
So far, we have not dealt with the quark sector. Since in the
above scenario we have assumed CP invariance at the 
Lagrangian level, the quark Yukawa couplings will 
also be real. It has been shown \cite{Bento:1990wv}
that the phase $\theta$ of $\langle S \rangle$ can still 
generate unsuppressed CP violation in the quark sector at 
low energies, provided one introduces at least one isosinglet
vector-like quark. These vector-like quarks play a 
r\^ ole analogous to the one played by right-handed
neutrinos in the leptonic sector, in the sense that they allow
the phase $\theta$ to be seen by the charged $W$ interactions
connecting standard quarks.
\section{Conclusions}
\setcounter{equation}{0}
We have studied the various sources of CP violation in the minimal
seesaw model, identifying both the CP violating phases 
and the weak-basis invariants which are associated to 
leptogenesis as well as those relevant for CP violation at 
low energies. We have addressed the question of whether it is possible 
to establish a connection between CP violation responsible
for leptogenesis and CP violation observable at low energies, for example
through neutrino oscillations. It was shown that, in general, 
such a connection between the two phenomena cannot be
established. However, we have described a class of models
where the phases which are responsible for leptogenesis are
also the ones that generate CP violation at low energies.
This situation naturally occurs in models with 
the feature of having,
in a weak-basis, the left-handed component of 
charged leptons aligned in flavour space with the 
left-handed neutrinos. In this class of models a direct
connection may in principle be established between
the size and sign of the observed baryon asymmetry obtained
through leptogenesis and CP violation observed at low energies.

\section*{Acknowledgements}

The authors thank T. Endoh, R. Gonzalez-Felipe, 
T. Onogi and A. Purwanto for useful discussions.
The work of TM was supported by a fellowship for
Japanese Scholar and Researcher abroad from the
Ministry of Education, Science and Culture of Japan.
The work of BMN was supported by Funda\c c\~ ao
para a Ci\^ encia e a Tecnologia (FCT) (Portugal) 
through fellowship SFRH/BD/995/2000; 
GCB, BMN and MNR received 
partial support  from FCT 
from Project CERN/P/FIS/40134/2000,
Project POCTI/36288/FIS/2000 and 
Project CERN/C/FIS/40139/2000.
GCB, TM and MNR thank the CERN Theory Division for
hospitality.

\end{document}